\documentclass[twoside,superscriptaddress,twocolumn,a4paper,floatfix,prl,aps,longbibliography]{revtex4-1}

\usepackage{xcolor}
\usepackage{graphicx}
\usepackage[utf8x]{inputenc}
\usepackage[T1]{fontenc}
\usepackage{siunitx}
\usepackage{amstext}
\usepackage{hyperref}
\usepackage{amsfonts}
\usepackage{amsmath}
\usepackage{tabularx}
\usepackage{filecontents}
\usepackage{soul}
\usepackage{dsfont}
\usepackage{mathtools}
\usepackage{makecell}

\makeatletter

\newcommand{\Rnum}[1]{\expandafter\@slowromancap\romannumeral #1@}

\begin{document}

\title{Experimental diagnostics of entanglement swapping by a collective entanglement test}

\author{Vojtěch Trávníček} \email{vojtech.travnicek@upol.cz}
\affiliation{RCPTM, Joint Laboratory of Optics of Palacký University and Institute of Physics of Czech Academy of Sciences, 17. listopadu 12, 771 46 Olomouc, Czech Republic}

\author{Karol Bartkiewicz} \email{bark@amu.edu.pl}
\affiliation{Faculty of Physics, Adam Mickiewicz University,
PL-61-614 Pozna\'n, Poland}
\affiliation{RCPTM, Joint Laboratory of Optics of Palacký University and Institute of Physics of Czech Academy of Sciences, 17. listopadu 12, 771 46 Olomouc, Czech Republic}

\author{Antonín Černoch} \email{antonin.cernoch@upol.cz}
\affiliation{RCPTM, Joint Laboratory of Optics of Palacký University and Institute of Physics of Czech Academy of Sciences, 17. listopadu 12, 771 46 Olomouc, Czech Republic}
   
\author{Karel Lemr}
\email{k.lemr@upol.cz}
\affiliation{RCPTM, Joint Laboratory of Optics of Palacký University and Institute of Physics of Czech Academy of Sciences, 17. listopadu 12, 771 46 Olomouc, Czech Republic}

\begin{abstract}
The paper reports on experimental diagnostics of entanglement swapping protocol by means of collective entanglement witness. Our approach is suitable to detect disturbances occurring in the preparation of quantum states, quantum communication channel and imperfect Bell-state projection. More specifically we demonstrate that our method can distinguish disturbances such as depolarization, phase-damping, amplitude-damping and imperfect Bell-state measurement by observing four probabilities and estimating collective entanglement witness. Since entanglement swapping is a key procedure for quantum repeaters, quantum relays, device-independent quantum communications or entanglement assisted error correction, this can aid in faster and practical resolution of quality-of-transmission related problems as our approach requires less measurements then other means of diagnostics.
\end{abstract}

\date{\today}

\maketitle
\paragraph*{Introduction.} The exchange of quantum information between parties connected through a quantum network \cite{Zeilinger2016,Jaeger2007,Bennett1992,Ralph1999,Croal2016} can be
inherently secure transmission of information \cite{Wootters1982,BEN84,Ekert1991,Renes2005} or provide improved transmission rate \cite{Harrow2004,Barreiro2008,Bartkiewicz2014}. Engeneering and diagnostics of multilevel quantum systems further show the possibility for improvement on the robustness and key rate of \emph{quantum communications} (QC) protocols \cite{Kues2017,Bavaresco2018}. Quantum teleportation \cite{Bouwmeester1997,Ma2012,Pirandola2015} appears to be one of the key protocols of QC providing an advantage over classical methods. In fact, early QC networks based on teleportation have already been proposed \cite{Pirandola2015,Loock2000,Castel2018,Barasiski2018} and realized experimentally \cite{Yonezawa2004,Barasinski2019}.

\begin{figure}[t]
\includegraphics[scale=0.99]{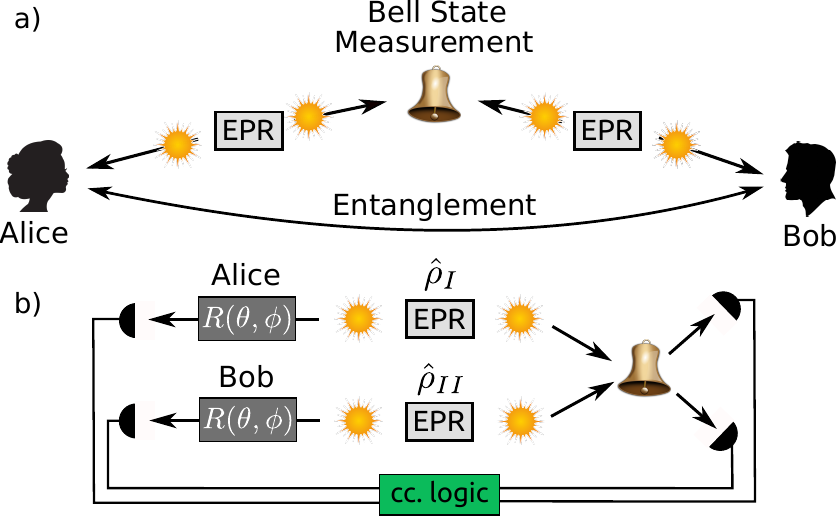}
\caption{\label{fig:scheme} (color online) a) Conceptual diagram of the ES protocol. Two entangled quantum states (e.g. two pairs of photons) are generated in EPR sources. One particle from each pair is subjected to a Bell-state measurement. This results in projecting the other two particles, which where sent to Alice and Bob, onto an entangled state. b) Conceptual scheme for measurement of a CEW. To witness entanglement of a general two-qubit state $\hat{\rho}_{I}$, a copy $\hat{\rho}_{II}$ is prepared. Similarly to ES, one particle from each pair is subjected to a Bell-state measurement, while the remaining two particles are subjected to a set of individual local projections $R(\theta,\phi)$. The CEW is then calculated from the rates of fourfold simultaneous detections observed for a specific set of local projections.}
\end{figure}

In an optical fiber the scaling of probability for a photonic qubit being absorbed, depolarized or
dephased grows exponentially with the length of the channel and remains to be the major obstacle to practical long-distance QC \cite{Briegel1998}. This does not only restrict feasible lengths of quantum channels, but also represents a security threat as the errors could be exploited for a potential attack on the communication protocol \cite{Bartkiewicz2013PRL, Bartkiewicz2017}. To combat these limitations, quantum repeaters and relays were proposed \cite{Briegel1998,Jacobs2002}. Although the working principles of quantum repeaters and relays somewhat differ, they both operate by splitting the communication channel into segments; therefore, lowering the error probability. At their core, quantum repeaters and relays apply the \emph{entanglement swapping} (ES) protocol \cite{Pan1998}. This involves teleportation of a quantum state of a particle that shares entanglement with at least one other particle. Thus, ES allows to establish entanglement between particles that have never interacted directly. By properly positioning the entanglement sources (EPR) and measurement device across the communication channel, one can distribute entanglement without physically sending the individual quantum-correlated information carriers through the entire channel (see Fig.\ref{fig:scheme}a). The ES is also applied in device-independent quantum communications or entanglement-assisted error correction \cite{Curty2011,Brun2006}.

In previous demonstrations of ES, quantum repeaters, and relays, the authors used various methods to demonstrate successful operation of their schemes. For instance, Li {\it et al.} \cite{Li2019} used quantum state tomography, Pan {\it et al.} \cite{Pan1998} and de Riedmatten {\it et al.} \cite{Riedmatten2004} observed interference visibility and Jennewein {\it et al.} \cite{Jennewein2001}, Zhao {\it et al.} \cite{Zhao2003} and Yuan {\it et al.} \cite{Yuan2008} tested Bell-inequality on the resulting state. In this paper, we present a practical method for diagnostics of ES by means of a \emph{collective entanglement witness} (CEW) \cite{Bartkiewicz2017a,Bartkiewicz2018,Horodecki2003,Bovino2005}. In particular, we adopt the {\it collectibility} witness originally proposed by Rudnicki {\it et al.} \cite{Rudnicki2011,Rudnicki2012}.  Our approach is preferable to diagnostics by other means because the number of measurement configurations is smaller (especially when compared to complete quantum state tomography \cite{Salles2008,Miranowicz2014,Bartkiewicz2016}). Moreover, the geometry of ES shares the layout of CEW (see Fig.~\ref{fig:scheme}), both protocols require simultaneous preparation of two copies of a given potentially entangled state and a Bell-state projection.

Collectibility was originally designed as a CEW for any approximately pure quantum state that requires a minimalistic set of four perfect measurements (i.e., perfect quantum channels and  Bell-state projection) on two copies of the state simultaneously. However, there is no difference between the results of (a) preparation of a mixed state propagating through a perfect channel and (b) generation of a pure state subsequently subjected to noise. Additionally, (c) imperfect Bell-state projection also influences, albeit differently, the measured photon statistics. As we show in Ref.~\cite{SM}, we can interpret the imperfect Bell-state projection as a type of a noisy quantum channel. The measurements constituting the CEW are able to detect even this source of imperfections. In the absence of any of the imperfections (a)-(c) the ES is successful and Alice and Bob must share an entangled state. 

We demonstrated this idea experimentally on a linear-optical platform, where we  constructed two independent EPR-state sources with an ES device linking them together. Qubits were encoded into polarization states of individual photons. We used polarizers and wave plates to implement errors occurring in three distinct quantum-information channels, i.e., (a) a depolarizing channel, (b) a phase-damping channel and (c) an amplitude-damping channel. In this experiment, both EPR pairs are subjected to the  identically prepared damping channels.

\begin{figure}[t]
\includegraphics[scale=1]{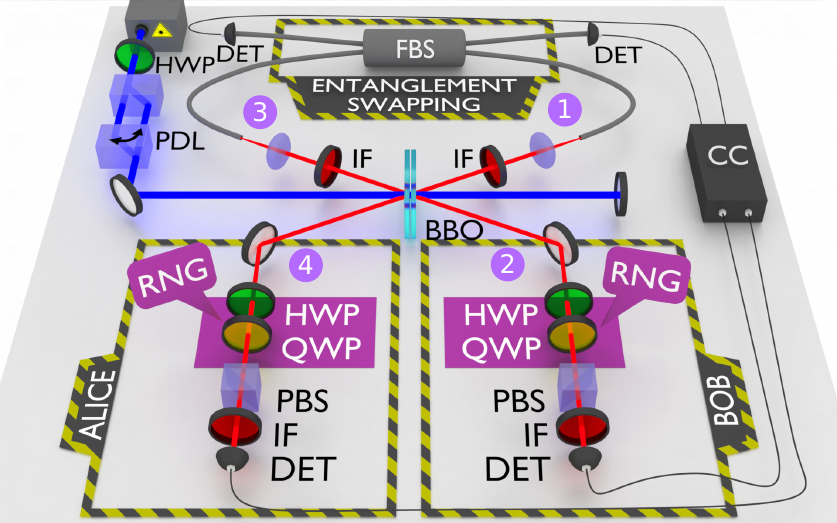}
\caption{\label{fig:Setup} (color online) Experimental setup for diagnostics of the ES protocol. HWP: half-wave plate, QWP: quarter-wave plate, PBS: polarization beam splitter, IF: \SI{10}{\nano\meter} interference filter, FBS: fiber beamsplitter, DET: single-photon detector, CC: coincidence counter unit.}
\end{figure}

\paragraph*{Experimental implementation.} In the experimental setup depicted in Fig.~\ref{fig:Setup}, a frequency-doubled \SI{413}{\nano\meter} femtosecond pulsed laser beam is used to pump spontaneous parametric down-conversion in a BBO crystal cascade \cite{Kwiat1999}. At first, the pump polarization is made diagonal. Next, the beam travels through a polarization dispersion line (PDL) to counter subsequent polarization dispersion of the BBO material. This laser beam impinges on the crystal cascade twice, i.e., after it passes the crystals for the first time, it gets reflected on a mirror and pumps the crystals in the opposite direction. On both occasions, with some probability, a pair of photons in the Bell state $|\Phi^+\rangle = \frac{1}{\sqrt{2}}\left(|HH\rangle + |VV\rangle\right)$ is generated, where $H$ and $V$ denote horizontally- and vertically-polarized photons, respectively. For a detailed characterization of this four-photon source see Ref.~\cite{Lemr2016}. While polarization of photons 2 and 4 is projected locally on four states selected using combinations of half- and quarter-wave plates followed by polarizers, the other two photons (1 and 3) are projected onto a singlet state by means of a balanced fiber coupler (FBS) and post-selected onto coincident detection at both its output ports.

The four specific settings of local projections sufficient to estimate collectibility \cite{Rudnicki2012} are: $|HH\rangle$, $|HV\rangle$, $|VV\rangle$ and $|++\rangle$, where letters indicate state projections on the two locally-projected photons respectively and $|+\rangle = \frac{1}{\sqrt{2}}\left(|H\rangle + |V\rangle\right)$. We denote $p_{XY}$ ($XY \in \lbrace HH,HV,VV,++\rbrace$) the probability that both locally-projected photons pass the projections conditioned on the other two photons being projected onto a singlet Bell state. However, due to nonremovable jitter between generation of the first and the second pair of photons, probability of two-photon overlap is decreased. These noninteracting photons are seen as noise, which can be estimated and subtracted from the genuine coincidences. In order to estimate the noise level both photon 1 and photon 3 were prepared in the same polarization state ($|H\rangle$) and the achivable Hong-Ou-Mandel bunching effect was measured conditioned on detection of photon 2 and 4 (used as heralds). From the visibility of Hong-Ou-Mandel interference the noise level caused by jitter was estimated (for more details see Ref.~\cite{Lemr2016}). Measuring the probabilities $p_{XY}$, collectibility is calculated using formula

\begin{equation}
\begin{split}
W(\hat{\rho}) = & \tfrac{1}{2}[\eta + p_{H}^{2}(1-2p_{HH}) + (1-p_{H})^{2}(1-2p_{VV})\\
                & + 2p_{H}(1-p_{H})(1-2p_{HV})-1],
\end{split}
\end{equation}
where 
$\eta = 16p_{H}(1-p_{H})\sqrt{p_{HH}p_{VV}} + 4p_{++},$
and $p_{H}$ is the probability of local projection of photon 1 or 3 onto horizontal polarization $|H\rangle$ independently of the singlet Bell-state projection.

For each measurement of the $p_{XY}$ probabilities there was 16 setup configurations of wave plates for photons 2 and 4. Wave plates used for local polarization projections were used simultaneously to introduce disturbances typical for a given type of erroneous quantum channel. As a result $16\times 60$ sequences of 4-fold coincidences were obtained. Depending on the simulated quantum channel a sequence was randomly with some bias selected and placed into a new array $cc_{XY}$ which in the end contained 60 randomly chosen sequences. The bias was dependent on the type of erroneous quantum channel and on the chance for error to occur. The probability $p_{XY}$  was then calculated by summation of $cc_{XY}$, normalization and correction on non-interacting photons. See Ref.~\cite{SM} for scheme of the experimental process. Here, we investigate experimentally noisy channels studied theoretically in the context of quantum teleportation in Ref.~\cite{Ozdemir2007}).

\begin{figure}
\includegraphics[scale=1]{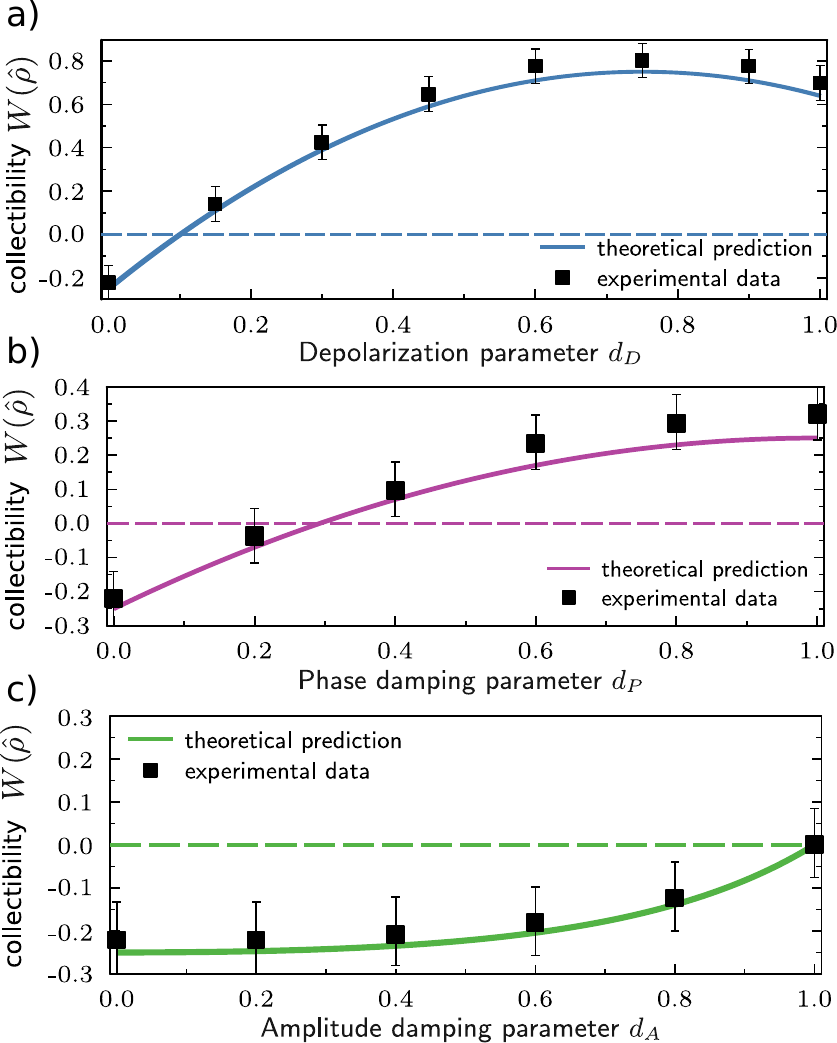}
\caption{\label{fig:experiment} (color online) Measured collectibility after the EPR pairs pass through: a) depolarizing channel b) phase-damping channel c) amplitude-damping channel  parameterized by parameter $d_D$, $d_P$ and $d_A$ respectively.}
\end{figure}

\paragraph*{Depolarizing channel.} Qubits transmitted through a depolarizing channel are randomly subjected to three types of transformations causing decoherence. These transformations are bit-flip, phase-flip and combination of bit-flip and phase-flip. It is the randomness and impossibility to predict these transformations that is the effective cause of errors. The action of such channel expressed as Kraus operators \cite{Nielsen2011} reads
\begin{equation}
\hat{E}_0 = \sqrt{1-d_{D}}\hat{I},\; \hat{E}_i = \sqrt{\frac{d_{D}}{3}}\hat{\sigma}_i \mathrm{\quad for\;} i \in \lbrace x,y,z\rbrace,
\end{equation}
where $d_{D}$ is the depolarization probability, $\hat{I}$ stands for the identity operator and $\sigma_{i}$ are Pauli matrices. When propagating through such channel, a Bell state is randomly transformed into one of the other three Bell states with equal probability $d_{D}/3$. To implement a depolarizing channel, we have been randomly switching the half-wave plates between two positions  $0^{\circ}$ and $45^{\circ}$ to achieve the bit-flip transformation and a quarter-wave plates between $0^{\circ}$ and $90^{\circ}$ to achieve the phase-flip transformation. Combined action of both wave plates implements the bit and phase-flip simultaneously. Using the procedure described in the previous paragraph we have been able to measure the collectibility of a Bell state propagating through a depolarizing channel for several values of the depolarization probability $d_{D}$. The observed collectibility and its theoretical prediction are depicted in Fig.~ \ref{fig:experiment} a). As expected, collectibility reaches its maximum value for $d_{D} = 3/4$, $W(\hat{\rho}) = 0.80 \pm 0.09$ (theoretical prediction: $W(\hat{\rho}) = 0.75$). This corresponds to a maximally depolarizing action causing the transmitted state to fully decohere to $\hat{\rho}_{D} = \hat{\mathds{1}}/4$. Meanwhile, in an ideal channel ($d_{D}=0$) the Bell state is propagating undisturbed which coincides with the value of collectibility being $W(\hat{\rho}) = -0.24 \pm 0.06$ (theoretical prediction: $W(\hat{\rho}) = -0.25$).

\paragraph*{Phase-damping channel.} The effect of phase damping causes decoherence between two basis qubit states without, however, causing any bit-flip transformation. Such channel is described by two Kraus operators
\begin{equation}
\hat{E}_0 = \sqrt{1-\frac{d_P}{2}}\hat{I},\qquad \hat{E}_1 = \sqrt{\frac{d_P}{2}}\hat{\sigma}_z,
\end{equation}
where $d_P$ is the dephasing probability. Similarly to the previous case, the phase-damping effect was implemented by randomly switching a quarter-wave plates between two positions: 0$^{\circ}$ and 90$^{\circ}$. The resulting collectibility as a function of $d_P$ is presented in Fig.~\ref{fig:experiment} b). Experimental value of collectibility at $d_{P} = 1$ reaches $W(\hat{\rho}) = 0.32 \pm 0.09$ (theoretical prediction: $W(\hat{\rho}) = 0.25$) as the Bell state propagating through this channel becomes $\hat{\rho}_{P} = \frac{1}{2}(|HH\rangle\langle HH|+|VV\rangle\langle VV|)$.


\paragraph*{Amplitude-damping channel.} Typically, amplitude damping causes lossy transmission of qubits through the channel. The overall losses are trivial to detect as they decrease the overall number of coincident detections. Apart from that, white (state-independent) losses do not change the collectibility because the measurement relies solely on successful four-photon detections. It is, therefore, more interesting to analyze state-dependent (polarization sensitive) losses that cause disturbance in superposition of horizontal and vertical polarizations of the state. We describe this channel by an effective matrix transformation
\begin{equation}
\hat{\rho} \rightarrow \hat{E}_A\hat{\rho}\hat{E}_A^\dagger,\qquad \hat{E}_A = 
\begin{pmatrix}
1 & 0\\
0 & \sqrt{1-d_A}
\end{pmatrix}.
\end{equation}
Here, in contrast to the above-described channels, the entangled state remains pure but its entanglement decreases. This corresponds to the Bell state being less entangled $\frac{1}{\sqrt{2-d_{A}}}(|HH\rangle + (1-d_{A})|VV\rangle)$ and eventually becoming separable $\hat{\rho}_{A} = |HH\rangle\langle HH|$ as $d_{A} \to 1$, where the value of collectibility reaches $W(\hat{\rho}) = -0.05 \pm 0.09$ (theoretical prediction: $W(\hat{\rho}) = 0$). Collectibility allows to capture this transition as shown in Fig.~\ref{fig:experiment}c. Note that the CEW for pure states can serve as an entanglement measure \cite{Rudnicki2011,Rudnicki2012}.


\begin{figure}
\includegraphics[scale=1]{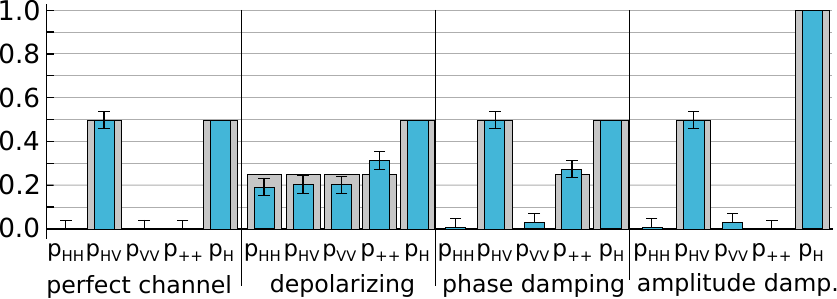}
\caption{\label{fig:Characteristic} (color online) Characteristic channel signatures allowing to identify type of errors from the individual measurements that constitute collectibility. Gray and blue bars represent theoretical predictions and experimentally obtained values respectively. The uncertainty of the unconditioned probability $p_H$ is negligible, therefore, is not visualized.}
\end{figure}

\paragraph*{Channel characteristics.} Measurement of collectibility is a powerful tool that allows to detect disturbance occurring in the channel. However, in order to promote our method even further, we have analyzed characteristic effects of the three types of erroneous channels. By detailed analysis of the individual probabilities used for collectibility calculation one can identify which type of damping is inflicted. Five probabilities are measured to calculate collectibility $p_{XY}$ for $XY \in \lbrace HH,HV,VV,++\rbrace,$ and $p_H$. We show experimental and theoretical values of these quantities for three tested channels and a reference perfect channel in Fig.~\ref{fig:Characteristic}. The exact results are then summarized in Tab.~\Rnum{2} of \cite{SM}. For a perfect channel the overall state of the system is
\begin{equation}
|\Phi^{+}\rangle|\Phi^{+}\rangle = \tfrac{1}{2}[(|HH\rangle + |VV\rangle)(|HH\rangle + |VV\rangle)],
\label{fifi}
\end{equation}
which after projecting the photons $1$ and $3$ onto a singlet state collapses also to a singlet state
\begin{equation}
|\Phi^{+}\rangle|\Phi^{+}\rangle\xrightarrow{|\psi^{-}_{13}\rangle\langle\psi^{-}_{13}|} |\psi^{-}_{24}\rangle.
\end{equation}
Hence, the only conditioned projection that we observe is the $|HV\rangle$ projection with probability $p_{HV}$ of 1/2. It follows from the Eq.~(\ref{fifi}) that the probability $p_{H}$ of unconditional projection $|H\rangle$ is also 1/2. In a fully depolarizing channel the state of the system becomes maximally mixed
\begin{equation}
\hat{\rho}_{D}\otimes \hat{\rho}_{D} = \hat{\mathds{1}}/16\xrightarrow{|\psi^{-}_{13}\rangle\langle\psi^{-}_{13}|} \hat{\mathds{1}}_{24}/4.
\end{equation}
Therefore, all of the conditional projections are equally likely with probabilities of $1/4.$ The probability $p_{H}$ of unconditional projection $|H\rangle$ stays at $1/2.$ Phase-damping transforms the initial Bell state into a $\hat{\rho}_{P}$. The final state of the photons $2$ and $4$ is then
\begin{equation}
\hat{\rho}_{P}\otimes\hat{\rho}_{P}\xrightarrow{|\psi^{-}_{13}\rangle\langle\psi^{-}_{13}|} \tfrac{1}{2}(|H_{2}V_{4}\rangle\langle H_{2}V_{4}|+|V_{2}H_{4}\rangle\langle V_{2}H_{4}|).
\end{equation}
The probability of observing a conditional $|HV\rangle$ projection is $1/2,$ however, due to the phase-flip transformation we also observe signal in $|++\rangle$ projection with probability $p_{++}$ of $1/4.$ The unconditional projection $|H\rangle$ happens with probability $1/2.$ In an amplitude-damping channel with attenuated vertical polarization the state of the photons $2$ and $4$ becomes
\begin{equation}
|\Phi^{+}\rangle|\Phi^{+}\rangle\xrightarrow{|\psi^{-}_{13}\rangle\langle\psi^{-}_{13}|} |\psi^{-}_{24}\rangle.
\end{equation}
As $d_{A} \to 1$ the probability of singlet projection ${|\psi^{-}_{13}\rangle\langle\psi^{-}_{13}|}$; however, tends to $0$ as the state becomes separable. In this limit, conditioned projection $|HV\rangle$ is observed with probability of $1/2,$ meanwhile, the probability $p_{H}$ of unconditioned projection $|H\rangle$ raises to $1.$



\paragraph*{Imperfect Bell-state measurement.}  To explicitly demonstrate that imperfect Bell-state projection also has a measurable effect and can be detected, we calculate the probabilities $p_{XY}$ without the compensation for non-interacting photons. This is specific to the linear-optical platform, where Bell-state projection is implemented by two-photon interference. The obtained values are $p_{HH} = 0.29$, $p_{HV} = 0.49$, $p_{VV} = 0.27$ and $p_{++} = 0.29$, with typical uncertainty of $0.03$. The resulting collectibility reads $W = 0.75\pm0.06$. These results prove that imperfect Bell-state projection by a balanced beam splitter is also detected by our method and additionally manifests a unique signature: the probability $p_{HV}$ should maintain a value of $0.5$ as in the case of a perfect channel, whereas the remaining probabilities should uniformly increase their values from $0$ (perfect Bell-state projection) up to $0.5$ (Bell-state projection replaced by completely non-interfering photons). Consequently, the imperfections in Bell-state projection can be distinguished from the channel imperfections. Our experimental results reflect the fact, that the Bell-state measurement was imperfect only to some degree.

\paragraph*{Conclusions.} We have reported on experimental diagnostics of entanglement swapping by utilizing four partial measurements applied for determining CEW (collectibility). We capitalize on the similarity between the geometry of ES protocol and the layout for measurement of CEW. This approach allows to detect disturbance in a channel by measuring four probabilities $p_{XY}$ and estimating the collectibility. This makes our approach a preferable method as the number of measurement configurations is lower than in other means of diagnostics. We have measured collectibility for three noisy channels: depolarizing channel, phase-damping channel and amplitude-damping channel. The obtained experimental data is in a good agreement with theoretical predictions. Additionally, by analysis of the measured $p_{XY}$ probabilities, we have determined which type of error is occurring in a given experimental setting for all three damping channels.  We have also experimentally demonstrated that our method is able to detect imperfections in Bell-state measurement and is capable of distinguishing them from previously-mentioned faulty channels. We believe that these results can contribute to the field of quantum communications and mainly represent a practical instrument for future deployment of quantum networks or engineering of complex multilevel quantum systems.

\paragraph*{Acknowledgement.} Authors thank Cesnet for providing data management services. Authors acknowledge financial support by the Czech Science Foundation under the project No. 20-17765S. KB also acknowledges the financial support of the Polish National Science Center under grant No. DEC-2019/34/A/ST2/00081. VT also acknowledges internal Palacky University grant IGA-PrF-2020-007. The authors also acknowledge the project No. CZ.02.1.01./0.0/0.0/16\textunderscore 019/0000754 of the Ministry of Education, Youth and Sports of the Czech Republic.


%

\end{document}